\documentclass[10pt,a4paper,oneside]{article}

\usepackage{xcolor}
\usepackage[margin=1in]{geometry}
\usepackage{amssymb}
\usepackage{amsmath}
\usepackage{amsthm}
\usepackage{multirow}
\usepackage{graphicx}
\usepackage{caption}
\usepackage[authoryear,round]{natbib}
\usepackage[colorlinks=true,linkcolor=blue,citecolor=green,urlcolor=magenta]{hyperref}
\usepackage[linesnumbered,ruled,vlined]{algorithm2e}
\usepackage{float}
\usepackage{tabularx}
\usepackage{setspace,booktabs,microtype,enumitem,subcaption,tikz}
\usetikzlibrary{arrows.meta}

\DeclareFontFamily{U}{mathx}{\hyphenchar\font45}
\DeclareFontShape{U}{mathx}{m}{n}{<->mathx10}{}
\DeclareSymbolFont{mathx}{U}{mathx}{m}{n}
\DeclareMathSymbol{\intop}{1}{mathx}{"B3}

\newcommand{\given}{\,|\,}
\newcommand{\norm}[1]{\Vert#1\Vert}

\newcommand{\parenthese}[1]{\left(#1\right)}
\newcommand{\cb}[1]{\left\{#1\right\}}
\newcommand{\gr}{\mathsf{G}}
\newcommand{\ver}{\mathsf{V}}
\newcommand{\edg}{\mathsf{E}}
\DeclareMathOperator{\tr}{tr}
\DeclareMathOperator*{\argmin}{arg\,min}
\DeclareMathOperator{\score}{score}
\newcommand{\diag}{\mathrm{diag}}
\newcommand{\subto}{\mathrm{subject\ to}}
\newcommand{\E}{\mathbb{E}}
\newcommand{\EE}{\mathbb{E}}
\newcommand{\R}{\mathbb{R}}
\newcommand{\RR}{\mathbb{R}}
\renewcommand{\P}{\mathbb{P}}
\newcommand{\Dags}{\mathbb{D}}
\newcommand{\B}{B} %
\newcommand{\trueB}{\B^0}
\newcommand{\A}{A} %
\newcommand{\p}{p} %
\newcommand{\EqClass}{\mathcal{E}(\P(\trueB))}
\newcommand{\MinEqClass}{\mathcal{E}_{\min}(\P(\trueB))}
\newcommand{\EqClassOrder}{\mathcal{T}(\P(\trueB))}
\newcommand{\MinEqClassOrder}{\mathcal{T}_{\min}(\P(\trueB))}
\newcommand{\bfx}{\mathbf{x}}
\newcommand{\bfX}{\mathbf{X}}
\newcommand{\PA}{\mathrm{PA}}
\newcommand{\pLD}{\rho_{\lambda,\delta}}

\newcommand{\myalg}{\textsc{GLM-DAGMA}}
\let\savedphi\phi
\let\phi\varphi
\let\varphi\savedphi

\newtheorem{theorem}{Theorem}
\newtheorem{lemma}{Lemma}

\theoremstyle{definition}
\newtheorem{definition}{Definition}
\newtheorem{remark}{Remark}

\setlist[enumerate,1]{label=(\alph*)}
\setlist[enumerate,2]{label=(\roman*)}

\allowdisplaybreaks[1]
\title{Nonasymptotic causal structure learning\\in generalized linear DAGs with continuous optimization}
\author{Chang Deng\qquad
  Inbeom Lee\qquad 
  Bryon Aragam}
\date{}

\begin{document}

\maketitle
{\renewcommand\thefootnote{}\footnotetext{Email: \texttt{\{changdeng,inbeom.lee,bryon\}@chicagobooth.edu}.}}

\begin{abstract}
We study the problem of learning directed acyclic graphs from observational data in nonlinear models and provide finite-sample, nonasymptotic statistical guarantees in these models. Although widely studied, statistical guarantees for gradient-based methods in nonlinear models are missing from the literature. Under a generalized linear model for each node, we analyze a global nonsmooth objective combining an exact acyclicity constraint with a quasi-minimax concave penalty, which can be optimized via standard gradient-based solvers. We establish finite-sample guarantees under this model, including the most widely used logistic and Poisson models. Experiments corroborate structure recovery accuracy and computational performance, including both standardized data and negative controls as baselines.
\end{abstract}

\noindent\textbf{Keywords:} causal discovery; continuous optimization; generalized linear models; nonasymptotic; finite-sample guarantees

\section{Introduction}\label{sec.intro}
Gradient-based approaches to model selection substantially simplify implementation and adoption of statistical procedures. Examples include the Lasso for variable selection \citep{tibshirani1996regression}, graphical Lasso for graphical models \citep{friedman2008sparse}, and nuclear norm minimization for matrix factorization \citep{recht2010guaranteed}. Recently, there has been increased interest in differentiable\footnote{Strictly speaking, these methods may not be differentiable since they may incorporate nonsmooth regularizers such as the $\ell_{1}$ and MCP. However, we use the common convention in this literature where ``differentiable'' refers to the differentiable acyclicity constraint and the use of gradient-based optimizers. See Section~\ref{sec.estimation}.} approaches to \emph{causal structure learning} based on differentiable acyclicity constraints, introduced in \citet{zheng2018dags}. Learning a directed acyclic graph (DAG) is a fundamental task that enables the discovery of underlying causal mechanisms \citep{scholkopf2021towards}, which can help explain unknown phenomena \citep{dominici2025causal}, aid in prediction accuracy, and boost out-of-distribution performance in downstream tasks \citep{lv2022causality,saengkyongam2023identifying,kong2024towards}. However, learning causal structures from observational data is notoriously challenging both from a theory and implementation perspective, namely due to non-identifiability and the combinatorial nature of the search space which leads to computational intractability in practice.

While detailed finite-sample properties are well understood in many settings for \emph{non-differentiable} methods, including the linear Gaussian \citep{shojaie2010penalized,vandeGeerBuhlmann2013l0DAG,gao2022optimal, xu2025asymptotically}, linear non-Gaussian \citep{loh2014high}, and non-linear \citep{park2018learning, park2019high,wang2024causal} settings, very little is known for \emph{differentiable} approaches based on continuous optimization. At the same time, differentiable methods \citep{zheng2018dags,yu2019dag,brouillard2020differentiable} are often used in practice due to their ease of implementation and convenient integration into existing differentiable machine learning (ML) pipelines. Despite progress on both fronts, explicit finite-sample theory for differentiable formulations remains comparatively limited, especially beyond linear Gaussian models. 

We study differentiable causal structure learning for DAGs with GLM conditional distributions, allowing nonlinear relationships. To this aim, we first cast causal structure learning as an estimation problem with a single global objective over all topological orders, where the objective is defined for a broad class of generalized linear models (GLMs) over a class of continuous DAG coefficient matrices. We refer to these as GLM-DAGs in the sequel. The GLM family is chosen as it is both general enough to permit nonlinear relationships, while also being concrete and explicit. Under this general framework, we give conditions under which the global minimizers of the objective are identifiable, and establish non-asymptotic finite-sample guarantees, improving both the scope and reliability of differentiable DAG learning. We then conduct extensive experiments to demonstrate a desirable balance of accuracy, speed, and ease of implementation, corroborating prior work tailored to the Gaussian case.

\paragraph{Overview.}
To present the main idea behind our results, we provide a high-level overview assuming minimal background on causal graphical models; formal development is deferred to Section~\ref{sec:prelim}. We are given $\bfX\in \R^{n\times p}$, a data matrix representing i.i.d. observations from a random vector $X=(X_1,\ldots,X_p)$ that is Markov to a DAG $\gr \in \Dags$, where $\Dags$ is the space of all directed acyclic graphs (DAGs) on $[\p]$. This implies that $X$ factorizes according to $\gr$, and we assume that the resulting conditional distributions follow a GLM with the canonical link: 
\begin{align}\label{equation:SEM}
\P(X_j\mid X_{\PA(j)}) = \exp \parenthese{X_j\eta_j - \A(\eta_j)}q(X_j), \qquad \eta_j = X^\top B_{\cdot j},\qquad j=1,\ldots, \p,
\end{align}
where $\PA(j)$ denotes the parent set of $X_j$. Since $\P(X_j\mid X_{\PA(j)})$ only depends on $X$ through its parents $X_{\PA(j)}$, it is implicit in \eqref{equation:SEM} that $B_{\cdot j}$ is also supported on $\PA(j)$. The $p$-dimensional column vectors $B_{\cdot j}$ make up the matrix $B=[B_{\cdot 1}\given\cdots\given B_{\cdot p}]\in\RR^{p\times p}$, the object of interest we wish to recover, as $B$ encodes the causal structure through its zero/non-zero pattern. 

Differentiable causal structure learning is related to \emph{score-based learning}, which refers to methods that aim to recover $B$ with a specified \emph{score function} --- typically based on a likelihood or posterior --- which is then optimized over the space of DAGs, $\Dags$. This is traditionally a combinatorial optimization problem that is difficult to solve due to the acyclicity constraint, but recent works starting with \citet{zheng2018dags} have introduced a smooth function $h(B)$ that allows this constraint to be enforced through a differentiable program, effectively casting the combinatorial problem into a continuous one. 
\begin{align}\label{eq:display.score}
    \begin{aligned}
    \min_{\gr}  \ \ \score(\gr;\bfX) \\
    \text{subject to}  \ \ \gr \in \Dags \\
    \text{(discrete,}\text{ combinatorial)}
    \end{aligned}
    \quad \iff \quad 
    \begin{aligned}
    \min_{B\in\RR^{p\times p}}  \ \ L(B;\bfX) \\
    \text{subject to} \ \ h(B) = 0 \\
    \text{(continuous,}\text{ differentiable)}
    \end{aligned}
\end{align}
The program \eqref{eq:display.score} shows differentiable causal structure learning alongside the general score based learning framework, with the constraint being specified as $h(B)$, a differentiable acyclicity constraint, and the score being specified as the penalized negative log-likelihood $L(B;\bfX):= \ell(B;\bfX) + \mathrm{pen}(B)$, where $\ell(\cdot)$, $\mathrm{pen}(\cdot)$ denote the negative log-likelihood and a sparsity penalty term, respectively. More details on notation and the specific components are deferred to Section \ref{sec:prelim}. In particular, we emphasize that this implicitly searches over all topological orders through a single continuous parameterization, which is not the case in general for traditional score-based learning methods.

The question then is: What are the statistical properties of the solutions to this global estimation problem? This includes an investigation into identifiability, consistency, and rates of convergence. 

\paragraph{Contributions.} 
Our main contributions are to provide explicit, non-asymptotic finite-sample rates for solutions to \eqref{eq:display.score} in the GLM-DAG setting. Along the way, this necessitates a detailed characterization of the identifiability of $B$ in this model, an issue that is well-known to plague DAG models in general. More precisely:

\begin{enumerate}
    \item We formulate non-linear causal structure learning from observational data as a single continuous global estimation problem over weighted adjacency matrices, combining a canonical GLM loss, an exact acyclicity constraint, and a nonconvex penalty that acts as a proxy for directed-edge counting.
    \item We prove a population identifiability result showing that, under minimal signal and local GLM curvature conditions, the global objective selects minimal-edge GLM-DAG representations. 
    \item We establish nonasymptotic order selection and coefficient estimation guarantees for every empirical global minimizer using empirical-process bounds on likelihood perturbations, localization arguments, and population separation conditions, without requiring a known topological order or skeleton.
    \item We evaluate our estimator across logistic, Poisson, and Gaussian DAGs, showing competitive accuracy-runtime tradeoffs and robust performance across sample sizes and negative control comparisons.
\end{enumerate}

\paragraph{Notation.} We let $\norm{X}_{\psi_1}$ denote the sub-exponential norm of random variable $X$ as defined in \citet{vershynin2018high}. For a vector $Z$, scalar $j$, and set $S$, we use $Z_j$ to denote its $j$-th component, and $Z_S$ to denote the subvector of components corresponding to the set $S$. For a matrix $W$, we denote the $j$-th column as $W_{\cdot j}$, and the $(i,j)$-th entry as $W_{i,j}$. For matrices with complex notation, we use square brackets for clarity, e.g. $[B(\pi)]_{[i,\pi(j)]}$ to denote the $(i, \pi(j))$-th element of the matrix $B(\pi)$, whereas for positive integer $p$, square brackets are used to denote the set of all positive integers up to $p$, i.e. $[p]:=\{1,...,p\}$. $\norm{W}_2$, $\norm{W}_F$, $\norm{W}_{\max}$ denote the spectral, Frobenius, and element-wise maximal norms, respectively, and $\norm{W}_0 := |\{(k,j):W_{k,j}\ne 0\}|$ denotes the number of non-zero entries in $W$. $\lambda_{\max}(W),~\lambda_{\min}(W), ~\lambda_{j}(W)$ are used to denote the largest, smallest, and $j$-th largest eigenvalue of a matrix $W$, respectively. For a random vector $X\in \RR^p$, its covariance matrix is denoted as $\Sigma_X \in \R^{p \times p}$. For any two sequences $a_n$ and $b_n$, we write $a_n \lesssim b_n$ if there exists some fixed positive constant $C$ such that $a_n \leq Cb_n$ for all $n$ (and vice versa). 

\subsection{Related work}%

\paragraph{GLM-DAGs / Non-linear SEMs.}
DAG learning is NP-hard in general \citep{chickering1996learning,chickering2004large}, and extensive statistical theory has been developed for linear Gaussian models \citep{shojaie2010penalized,vandeGeerBuhlmann2013l0DAG,loh2014high,aragam2019globally,li2023inference,chen2024discovery,xu2025integer}. Beyond Gaussian models, existing work studies multinomial-logit DAGs with interventions \citep{gu2019penalized}, instrumental variable methods under unmeasured confounding \citep{wang2024causal}, and distributed learning \citep{ye2024federated}. Other approaches exploit Poisson overdispersion \citep{park2015learning,park2019high}, quadratic variance functions \citep{park2018learning}, additive-noise assumptions \citep{park2020identifiability}, and hypergeometric conditional distributions \citep{park2019identifiability}. These approaches often rely on a discrete structure search or specialized, model-specific recovery procedures. Our focus is a single continuous objective over all candidate orders, without a separately obtained/supplied topological order or skeleton.

\paragraph{Differentiable structure learning.}
Differentiable methods encode acyclicity through smooth constraints on weighted adjacency matrices, enabling gradient-based optimization \citep{zheng2018dags,zheng2020learning,bello2022dagma,nazaret2023stable}. Extensions cover nonparametric and semiparametric models \citep{zheng2020learning}, neural-network models such as GraN-DAG and DAG-GNN \citep{lachapelle2019gradient,yu2019dag}, interventions \citep{brouillard2020differentiable,faria2022differentiable}, time series \citep{pamfil2020dynotears}, and unmeasured confounding \citep{bhattacharya2021differentiable}. Alongside these developments, theoretical and empirical studies examine when continuous approaches succeed or fail \citep{reisach2021beware,xu2022sparse,ng2024structure} and their optimization properties \citep{wei2020dags,ng2022convergence,deng2023optimizing}. More recently, \citet{deng2024markov,deng2025differentiable} study differentiable programs with smooth nonconvex penalties in general parametric families. While existing theory largely concerns population identifiability, asymptotics, or optimization, we establish finite-sample guarantees that go beyond Gaussian models.

\section{Model assumptions and identifiability}
\label{sec:prelim}

\subsection{Preliminaries}

We let $\gr = (\ver,\edg)$ denote a directed graph on $\p$ nodes, with vertex set $\ver = [\p]:=\{1,\ldots,\p\}$ and edge set $\edg \subset \ver\times \ver$, where $(i,j)\in \edg$ indicates the presence of a directed edge from node $i$ to node $j$. Let the parent set of node $j$ be $\PA(j) = \{i: (i,j)\in \edg\}$.  We associate each node $j\in \ver$ to a random variable $X_j$, express the vector as $X = (X_1,\ldots,X_\p)$, and express the corresponding data as $x = (x_1,\ldots,x_\p)^\top$, with the distribution of $X$ denoted as $\P(X)$. For data over multiple observations, we use the notation $\mathbf{X} = (\mathbf{X}_{\cdot 1},\ldots,\mathbf{X}_{\cdot p}) = (\bfx_{\cdot 1},\ldots,\bfx_{\cdot n})^\top \in \mathbb{R}^{n\times \p}$, where $\bfx_{\cdot i}\in\mathbb{R}^p$ denotes the $i$-th observation and $\mathbf{X}_{\cdot j}\in\mathbb{R}^n$ denotes the $j$-th variable. We also have a weighted adjacency matrix $\B = [B_{\cdot 1}\given\cdots\given B_{\cdot p}] \in \R^{\p \times \p}$ where $B_{k,j}\ne 0\Longleftrightarrow X_k\rightarrow X_j$ follows. Given a weighted adjacency matrix $\B$, define the induced directed graph $\gr(\B)$ as
\begin{align*}
    \gr(\B) := (\ver, \edg(\B)),\quad \ver = [p], \quad \edg(\B) := \cb{(k,j)\in[p]\times [p]:k\ne j, B_{k,j}\ne 0},
\end{align*}
and let $\Dags$ denote the space of all directed acyclic graphs (DAGs) on $[\p]$. In DAGs, the ordering of the nodes is of particular importance. We give a formal definition below.

\begin{definition}[Topological order]
A topological order is a permutation $\pi$ of $V$ such that
\[
X_{\pi(i)} \to X_{\pi(j)} \Longrightarrow i<j.
\]
\end{definition}

In other words, the permutation orders the nodes so that every directed edge points from an earlier to a later position. A directed graph is acyclic if and only if at least one such ordering exists. We write $B\sim \pi$ if the support of $B$ is consistent with the topological order $\pi$.

\subsection{Model}
As mentioned in the overview in Section \ref{sec.intro}, we consider structural equation models (SEMs) whose conditional distributions belong to an exponential family with the canonical link and factorize according to a DAG. More precisely, let $\P$ be a distribution that is Markov to a DAG $\gr$ and that satisfies \eqref{equation:SEM} for each $j$. In \eqref{equation:SEM}, $\eta_j$ denotes the natural parameter and $\A$ denotes the cumulant function, and $q$ denotes the base-measure factor of the chosen exponential family. We refer to this model as a GLM-DAG. Under this model, the joint distribution factorizes as
\begin{align}
    \P(X;\B)=\prod_{j=1}^\p \P(X_j\mid X_{\PA(j)}), \label{eq.chain}
\end{align}
where each conditional distribution is given by \eqref{equation:SEM}. We write $\P(B)$ when the context is clear. We assume that the data are generated from a true parameter $\trueB$, with corresponding DAG $\gr^0$, so that the population distribution is $\P(\trueB)$. Accordingly, we define the per-sample node-wise negative log-likelihood and its population counterpart as 
\begin{align}
    \ell_{j,n}(B_{\cdot j}; x) ~&:=~ - x_j\eta_j + \A(\eta_j) ~=~ -x_j\big(x^\top B_{\cdot j}\big) +\A\big(x^\top B_{\cdot j}\big), \label{eq.loss}\\
    \ell_{j}(B_{\cdot j}) ~&:=~ \E \big[\ell_{j,n}(B_{\cdot j}; x)\big], \label{eq.loss.2}    
\end{align}
respectively, where the decomposition in \eqref{eq.loss} follows from \eqref{equation:SEM}. We subsequently define the per-sample loss, empirical risk, and population risk as 
\begin{align*}
    \ell_n(\B;x)
   := \sum_{j=1}^\p \ell_{j,n}(\B_{\cdot j};x)
     , \qquad
    \ell_n(\B;\bfX)
    := \frac{1}{n}\sum_{i=1}^n \ell_n(\B;\bfx_{\cdot i}), \qquad
    \ell(\B)
    := \E\bigl[\ell_{n}(\B;X)\bigr], \nonumber
\end{align*}
respectively. We remark that the subscript $j$ is explicitly written in the node-wise loss functions $\ell_{j,n}(\cdot), \ell_j(\cdot)$ to clearly specify the response node even in situations where the argument vector is a generic $b \in \RR^p$. Again, it is implied that the arguments satisfy $B_{j, j}=0$ (or for generic $b$, that $b_j=0$). Also, we leave $n$ explicitly in the subscript for the per-sample loss $\ell_{n}(B;x)$ to more clearly distinguish it from the population version, $\ell(B)$.

To further illustrate GLM-DAGs, we include a few concrete examples; see Section \ref{sec.experiments} for empirical results on these models.

\paragraph{Logistic model.}
The logistic SEM with the logistic link is suitable for binary data. For each node $j\in [p]$,
\begin{align}
    X_j \mid X_{\PA(j)} \sim \mathrm{Bernoulli}\bigl(\sigma(B_{\cdot j}^\top X)\bigr), \label{eq.logistic.1}
\end{align}
where $\sigma(t)=1/(1+\exp(-t))$ is the sigmoid function, and $B_{\cdot j}$ is supported on the parent set $\PA(j)$. The empirical risk (up to an additive constant) is
\begin{align}
    \ell_n(B;\bfX)
    \;=\;
    \frac{1}{n}\sum_{i=1}^n \sum_{j=1}^p
    \left[
        \log\!\Big(1+\exp(\eta_{ij})\Big)
        - \bfX_{i,j}\,\eta_{ij}
    \right],
    \qquad
    \eta_{ij}: = \bfx_{\cdot i}^\top B_{\cdot j}.
    \label{eq:logistic_nll}
\end{align}

\paragraph{Poisson model.}
The Poisson SEM with the log-link is suitable for count data. For each node $j\in[p]$,
\begin{align}
  X_j \mid X_{\PA(j)} \sim \mathrm{Poisson}\big( \exp(B_{\cdot j}^\top X)\big), \label{eq.poisson.1}
\end{align}
where $B_{\cdot j}$ is supported on the parent set $\PA(j)$. The empirical risk (up to an additive constant) is
\begin{align}
    \ell_n(B;\bfX)
    \;=\;
    \frac{1}{n}\sum_{i=1}^n \sum_{j=1}^p
    \left[
        \exp(\eta_{ij}) - \bfX_{i,j}\,\eta_{ij}
    \right],
    \qquad
    \eta_{ij}: = \bfx_{\cdot i}^\top B_{\cdot j}.
    \label{eq:poisson_nll}
\end{align}

\paragraph{Linear Gaussian model.} 
The linear Gaussian model incorporates the identity link, and is suitable for continuous real-valued data. The data is generated according to the linear SEM
\begin{align}
   X = B^\top X +N \label{eq.Gaussian.1}
\end{align}
where $B_{\cdot j}$ is supported on the parent set $\PA(j)$ for all $j$, and $N\sim\mathcal{N}(0,\Omega)$ with $\Omega=\diag(\sigma_1^2,\ldots,\sigma_p^2)$. The empirical risk (up to an additive constant independent of $B$ and $\Omega$) can be written as
\begin{align}
  \ell_n(B,\Omega;\bfX)
  \;=\;
  \frac{1}{2n}\sum_{j=1}^p
  \left[
    \frac{\norm{\bfX_{\cdot j}-\bfX B_{\cdot j}}_2^2}{\sigma_j^2}
    + n\log \sigma_j^2
  \right].
  \label{eq:gaussian_nll}
\end{align}
 
\subsection{Sparsity and equivalence classes}\label{subsec:background}
Observational data alone cannot distinguish the ``true'' DAG parameter $\B^0$ since different topological orders $\pi$ can induce the same observational distribution $\P(X;\B^0)$.
A natural way to address this ambiguity is to select the \emph{sparsest} solution among all compatible DAG parameters, thereby favoring the simplest network representation of the distribution \citep{raskutti2018learning}. We pursue this through a penalized loss.

In order to select the sparsest solution among the compatible DAG parameters, a common approach is to impose an $\ell_0$ penalty \citep{vandeGeerBuhlmann2013l0DAG}, which directly penalizes the number of nonzero entries in $\B$. However, this leads to a combinatorial program that is incompatible with the \textit{differentiable} framework we are pursuing. A common convex surrogate is the $\ell_1$ penalty, which is convex and (sub)differentiable, but it penalizes the \emph{magnitude} of edge weights rather than directly counting the number of edges. Instead, we use a nonconvex penalty inspired by the the minimax concave penalty \citep[MCP][]{zhang2010nearly}, which better approximates the $\ell_0$ penalty. We adopt the following \emph{quasi-MCP} penalty: Given $\lambda>0$ and $\delta>0$, define 
\begin{align}\label{eq:quasi_MCP}
      \pLD(z)
    = \begin{cases}
      \lambda \bigl(|z| - \frac{z^2}{2\delta}\bigr),
      & |z|<\delta,\\[8pt]
      \displaystyle\frac{\lambda\,\delta}{2},
      & |z|\ge\delta,
    \end{cases}
     \quad \pLD(b)=\sum_{j}\pLD(b_{j}), 
    \quad 
    \pLD(B)=\sum_{k,j}\pLD(\B_{k,j}),
  \end{align}
for scalars ($z$), vectors ($b$), and matrices ($B$), respectively. This naturally leads to the following penalized sample-level and population loss functions:
\begin{align}
  L_n(\B;\bfX, \lambda,\delta)
  \;&=\;
  \ell_n(\B;\bfX) + \pLD(\B), \label{eq:sample_obj} \\
  L(\B;\lambda,\delta)
  \;&=\;
  \ell(\B) + \pLD(\B), \label{eq:population_obj}
\end{align}
along with the node-specific counterparts defined correspondingly, i.e. $L_{j,n}(B_{\cdot j}; \bfX, \lambda,\delta)$ defined with $\ell_{j,n}(B_{\cdot j})$, $\rho_{\lambda,\delta}(B_{\cdot j})$, and $L_{j}(B_{\cdot j}; \lambda,\delta)$ defined with $\ell_{j}(B_{\cdot j})$, $\rho_{\lambda,\delta}(B_{\cdot j})$.

We are now ready to formally characterize GLM-DAG learning as a constrained optimization problem. Given $n$ i.i.d. observations $\bfX\in\mathbb{R}^{n\times p}$ sampled from $\P(\trueB)$, we define the global finite-sample estimator $\widehat \B$ and population target $\B^*$ as 
\begin{align}
\begin{split}
\widehat B ~&:=~\argmin_{\B\in \R^{\p\times \p}}~   L_n(\B;\bfX,\lambda,\delta) ~~~
\text{subject to} ~~ \gr(\B)\in \Dags,
\end{split}\label{opt:dag_learn_finite}\\
\begin{split}
\B^*&:=~\argmin_{\B\in \R^{\p\times \p}}~  L(\B;\lambda,\delta) ~~~~~~~~~ 
\text{subject to} ~~  \gr(\B)\in \Dags,
\end{split}\label{opt:dag_learn_population}
\end{align}
respectively, where the constraint is that the graph corresponding to $\B$ is a DAG. The connection with an explicit differentiable acyclicity constraint is given in \eqref{opt:notears_finite} and will be motivated in Section \ref{sec.estimation}. These estimators can be formally collected into the following classes: 
\begin{align}
    \mathcal{M}_{n, \lambda,\delta} &= \bigl\{ \widehat{\B}: \widehat{\B} \text{ is a minimizer defined in } \eqref{opt:dag_learn_finite}\bigr\}, \label{eq.estimator.B}
    \\
    \mathcal{M}_{\infty, \lambda,\delta} &= \bigl\{ \B^*: \B^* \text{ is a minimizer defined in } \eqref{opt:dag_learn_population}\bigr\},
\end{align}
where we can also denote the topological order that the estimator $\widehat B$ follows in \eqref{eq.estimator.B} as $\widehat \pi$. While a single $\widehat B$ may correspond to multiple topological orders, we still use the notation $\widehat \pi$ to denote the order since any one of them can equivalently represent $\widehat B$. The equivalence class $\EqClass$ given in \eqref{eq:equivalence.class} is the collection of all the $\B$ that can generate the distribution $\P(B^0)$ through the SEM \eqref{equation:SEM}: 
\begin{align}\label{eq:equivalence.class}
        \EqClass = \cb{\B: \P(B^0) = \P(B)  }.
\end{align}

We are interested in the sparsest graph that can generate the distribution $\P(B^0)$:
\begin{definition}
    $\B$ is called a minimal-edge $\mathcal I$-map in the equivalence class $\EqClass$ if $\B\in \EqClass$ and $\norm{B}_0\leq \norm{\widetilde{\B}}_0 $ for all $\widetilde{\B}\in \EqClass$. 
\end{definition}
The minimal equivalence class is then defined as the collection of all minimal-edge $\mathcal I$-maps in the equivalence class $\EqClass$, i.e.
\begin{align}
    \MinEqClass = \cb{\B: \B \text{ is a minimal-edge $\mathcal I$-map in } \EqClass  }.
\end{align}
We collect all topological orders that are consistent with elements in $\EqClass$ and $\MinEqClass$ and denote them as
\begin{align}
    \EqClassOrder &= \cb{\pi:\exists \B \in \EqClass ~\mathrm{s.t.}~ B\sim \pi},\\
    \MinEqClassOrder &= \cb{\pi:\exists \B \in \MinEqClass ~\mathrm{s.t.}~ B\sim \pi}.
\end{align}
Although the true parameter $\B^0$ need not lie in the minimal equivalence class $\MinEqClass$, our target is the class of sparsest GLM-DAG representations of its observational distribution. For notational convenience, we henceforth use $\B^0 $ to denote an arbitrary representation of $\MinEqClass$.

Lastly, we define the minimal nonzero weight in the equivalence class $\EqClass$, which can be regarded as the ``minimal signal'' (using the convention $\min\varnothing=+\infty$):
\begin{align} \label{eq:tau}
    \tau = \min_{\B\in \EqClass}\min_{\{(k,j):|\B_{k,j}|\ne 0\} } |\B_{k,j}|.
\end{align}

\subsection{Identifiability over topological orders}\label{subsec:identifiability}

As alluded to before, identifiability is known to be a thorny issue in DAG models, as there could be multiple topological orders that are consistent with the observational data. Discussing identifiability in the more general nonlinear setting would only make things more nontrivial. Thus, in moving towards a theoretically justified \emph{global} estimator, a proper identifiability result for the GLM-DAG that takes all topological orders into account needs to be established. To compare representations across orders, let $S_j^\pi:=\{\pi(1),\ldots,\pi(j-1)\}$ denote the candidate parent set of node $\pi(j)$. For any order $\pi$, we consider the non-penalized population problem
\begin{align}\label{opt:notears_population_pi}
\begin{split}
  \min_{\B\in\R^{p\times p}} \quad & \ell(\B) \\
  \subto \quad & \B \sim \pi,
\end{split}
\end{align}
where the optimal solution is denoted by $\widetilde\B(\pi)$. Since this problem decomposes into $p$ node-wise problems, we define the corresponding node-wise population negative log-likelihood losses that respect the topological order $\pi$ and take in compatible-size argument vectors $b \in \RR^{j-1}$ as: 
\begin{align}
    \ell_{j}^{\pi}(b)
    &:= \EE \big[\A(\eta^\pi_{j}(b)) - x_{\pi(j)}\eta_{j}^{\pi}(b)\big], ~~~~\eta_{j}^{\pi}(b):=x_{S_j^{\pi}}^\top b, \label{eq:pop.log.lik.pi_j}
\end{align}
and write $\ell^\pi_{j,n}(b)$ for the empirical counterpart whose expectation equals $\ell^\pi_j(b)$. We then define the node-wise problems to be 
\begin{align}\label{opt:notears_population_pi_j}
  \min_{b \in \RR^{j-1}} \quad & \ell_{j}^{\pi}(b). 
\end{align}
For notational simplicity, the minimizer of \eqref{opt:notears_population_pi_j} can be expressed as $\widetilde{b}_j^{\pi}:= [\widetilde\B(\pi)]_{[S_j^{\pi},\pi(j)]} \in \RR^{j-1}$ and the corresponding optimal natural parameter as $\widetilde\eta_j^{\pi}:=x^\top_{S_j^{\pi}}\widetilde{b}_j^{\pi} \in \RR$. We add that \eqref{eq:pop.log.lik.pi_j} is the version of $\ell_j(B_{\cdot j})$ in \eqref{eq.loss.2} that has the dependence on the topological order $\pi$ explicitly baked in.

We will use the following conditions in the identifiability and the finite sample theory. 

\begin{enumerate}[label=(A\arabic*)]
\item\label{assum.subex}
The covariance satisfies $\lambda_{\max}(\Sigma_X)\leq\sigma_{X,\max}$. In addition, one of the following holds:
\begin{enumerate}[label=(\alph*)]
\item \emph{Bounded case:} $\norm{X}_{\max}\leq C_{\max}$ almost surely for a fixed $C_{\max}\geq1$.

\item \emph{Exponential tail bounds:} For $A(\eta)=e^\eta$ or $A(\eta)=\eta^2/2$, there exists a fixed $C_{\mathrm{tail}}\geq1$ such that, for every $u\geq0$,
\begin{equation}\label{eq.sample.tail}
\begin{aligned}
 \mathbb P\!\left(
 \max\left\{\norm{X}_{\max},
       \max_{\pi,j}|A'(\widetilde\eta_j^\pi)|\right\}>u\right)
 \leq ep\cdot \exp\!\left(-\frac{u}{C_{\mathrm{tail}}}\right).
\end{aligned}
\end{equation}
\end{enumerate}
\end{enumerate}

\begin{enumerate}[label=(A\arabic*),start=2]
\item\label{assum_glm} For every $\pi$, the population minimizer $\widetilde B(\pi)$ exists as a finite matrix. For every $\pi$ and $j\geq2$, the Fisher information matrix
\begin{align*}
 I_{\pi,j}:=\mathbb E\!\left[
 A''(\widetilde\eta_j^\pi)
 X_{S_j^\pi}X_{S_j^\pi}^{\top}\right]
\end{align*}
is finite and satisfies $\lambda_{\min}(I_{\pi,j})\geq\kappa$ for a fixed $\kappa>0$. In addition, one of the following holds:
\begin{enumerate}[label=(\alph*)]
\item \emph{Bounded case:} There exist fixed constants $C_{\mathrm{reg}},C_\eta>0$ such that $|X_{S_j^\pi}^{\top}b|\leq C_\eta$ almost surely whenever $\norm{b-\widetilde b_j^\pi}_2\leq C_{\mathrm{reg}}/\sqrt p$, and the aforementioned $\kappa>0$ satisfies $\inf_{|\eta|\leq C_\eta}A''(\eta)\geq\kappa$.

\item \emph{Exponential tail bounds:} Assume, for every $v\in\mathbb R^{j-1}$ and the aforementioned $\kappa>0$ that
\begin{align*}
 \mathbb E[(X_{S_j^\pi}^{\top}v)^2]
 \leq\frac1\kappa v^\top I_{\pi,j}v.
\end{align*}
\end{enumerate}
\end{enumerate}

\begin{remark}[Exponential tails and logarithmic sample bounds]\label{rem.sample.tails} Assumption \ref{assum.subex}(a) covers logistic and fixed-trial binomial DAGs, while \ref{assum.subex}(b) controls the exponential tails of the observations and working means, applicable to Poisson and canonical Gaussian DAGs. In particular, a union bound under \ref{assum.subex}(b) gives, for sufficiently large fixed $C_{\max}$,
\begin{equation}\label{eq.sample.bounds}
\begin{aligned}
 \max_{1\leq i\leq n}\norm{X_{(i)}}_{\max}
 ~&\leq~ C_{\max}\log(np),\\
 \max_{i,\pi,j}\left|A'\!\left(
 [X_{(i)}]_{S_j^\pi}^{\top}\widetilde b_j^\pi\right)\right|
 ~&\leq~ C_{\max}\log(np),
\end{aligned}
\end{equation}
simultaneously with probability at least $1-1/(4pn)$. These bounds have the same logarithmic scaling as \citet[Assumption 3]{park2019high}.
\end{remark}

\begin{remark}\label{rem.incompatible}
GLM conditional families need not be closed under reordering \citep{heckerman1998tutorial,loh2014high}. For incompatible orders, $\widetilde B(\pi)$ minimizes a working GLM risk, and \ref{assum_glm} imposes regularity on these targets, as in misspecified likelihood and GLM model selection theory \citep{fan2013tuning,lv2014model,demirkaya2022large,ye2024federated}.
\end{remark}

In light of Remark \ref{rem.incompatible}, we incorporate potential incompatible topological orders with \eqref{gap}. $g_{\mathrm{gap}}$ quantifies the log-likelihood gap between compatible and incompatible orders. For an incompatible order, $\ell(\widetilde B(\pi))-\ell(B^0)>0$ holds by the KL identity, and since there are finitely many orders, $g_{\mathrm{gap}}>0$ holds whenever incompatible orders exist. 
\begin{equation}\label{gap}
 g_{\mathrm{gap}}:=
 \begin{cases}
 \displaystyle
 \min_{\pi\notin\mathcal T(\mathbb{P}(B^0))}
 \bigl\{\ell(\widetilde B(\pi))-\ell(B^0)\bigr\},
 &\text{if there exists }\pi\notin\mathcal{T}(\mathbb{P}(B^0))\\[2mm]
 +\infty,&\text{otherwise}.
 \end{cases}
\end{equation}
We are now ready to present the main identifiability result that recovers the minimal structure among all possible topological orders.

\begin{theorem}[Identifiability]\label{thm:identifiability}
Given that the data are generated by SEM \eqref{equation:SEM}, and that \ref{assum.subex}-\ref{assum_glm} hold, let $s_0:=\|B^0\|_0$ be the number of edges in the sparsest representations. 

If $s_0 =0$, then for every $\lambda>0$ and $\delta>0$,  $\MinEqClass = \mathcal{M}_{\infty, \lambda,\delta} = \{0\}$.

If $s_0 \geq1$, then for any $\Delta>0$, define 
\begin{align}
\delta_0 ~=~ \tau/(1+\Delta) \qquad 
\lambda_0  ~=~   \min\cb{ \frac{\kappa}{2}\min\left( \frac{  r_0 \Delta}{\sqrt{s_0}},\frac{\Delta^2\tau}{1+\Delta}\right),\frac{2g_{\mathrm{gap}}}{s_0\delta_0}}, \label{eq.lambda_0}
\end{align}
where $\tau$ is defined in \eqref{eq:tau}, $g_{\mathrm{gap}}$ is defined in \eqref{gap} and $r_0$ is a local curvature radius specified in \eqref{eq.r_0_main}. Then, for every $0<\delta\leq \delta_0$ and $0<\lambda\leq \lambda_0$, $\MinEqClass = \mathcal{M}_{\infty, \lambda,\delta}$, i.e. global optimizers of \eqref{opt:dag_learn_population} are exactly the minimal-edge $\mathcal{I}$-maps of the data distribution $\P(B^0)$.
\end{theorem}
Under the additional SMR assumption, all population global minimizers induce Markov-equivalent DAGs.

\section{Estimation and implementation}\label{sec.estimation}

\subsection{Differentiable acyclicity constraints}
While theoretical identifiability can be discussed without specifying \emph{how} acyclicity is enforced in \eqref{opt:dag_learn_finite} and \eqref{opt:dag_learn_population}, this is well-known to be challenging to implement in practice because the constraint $\gr(\B)\in\Dags$ is inherently combinatorial. This motivates replacing the discrete constraint by an exact differentiable characterization of acyclicity, introduced in \citet{zheng2018dags}, enabling continuous optimization over $\B$ while still enforcing $\gr(\B)\in\Dags$. Thus, a central idea in differentiable DAG learning is to construct a smooth function $h:\R^{\p\times \p}\to \R_{\geq 0}$ such that 
\begin{align}
\label{eq:smooth_cons}
    \gr(\B)\in\Dags \quad \Longleftrightarrow \quad h(\B)=0.
\end{align}
This provides an exact smooth characterization based on the matrix exponential:
\begin{align}\label{eq:acyc_expm}
h_{\exp}(\B):=\tr\left(\exp(\B\circ \B)\right)-p,
\qquad
h_{\exp}(\B)=0 \ \Longleftrightarrow \gr(\B)\in\Dags,
\end{align}
where $\circ$ denotes the Hadamard product. Intuitively, $\exp(\B\circ\B):=\sum_{k=0}^\infty (\B\circ\B)^k/k!$ aggregates information from all matrix powers. Recall that the $(i,i)$-th entry of $(\B\circ\B)^k$ sums squared weights of all directed walks of length $k$ that start and end at node $i$. Therefore, taking the trace counts the total weight of closed directed walks. If the graph is acyclic, there are no directed cycles and hence only $k=0$ contributes to the value of the trace, which is $p$. Conversely, any closed directed walk would make the trace value strictly greater than $p$. Using \eqref{eq:acyc_expm}, the sample-level problem \eqref{opt:dag_learn_finite} can be written as the continuous constrained optimization problem
\begin{align}\label{opt:notears_finite}
    \min_{\B\in\R^{p\times p}} \ L_n(\B;\bfX,\lambda,\delta)\quad \text{subject to }h_{\exp}(\B) = 0.
\end{align}
Other exact differentiable acyclicity constraints have been proposed \citep{bello2022dagma,yu2019dag,nazaret2023stable}, mainly to improve numerical behavior. Our theory applies to these methods as well since our results are not tied to the choice of constraint, as long as \eqref{eq:smooth_cons} is satisfied.

In practice, \eqref{opt:notears_finite} is commonly solved by reformulating the equality constraint $h_{\exp}(\B)=0$ as a sequence of unconstrained optimization problems using either penalty methods or augmented Lagrangian methods \citep{zheng2018dags,bello2022dagma}. These approaches optimize an objective that augments $L_n(\B;\bfX,\lambda,\delta)$ with terms that penalize violations of the acyclicity constraint, for example through a quadratic penalty $\frac{\nu}{2}h(\B)^2$. Starting from an initial iterate, one repeatedly minimizes the resulting penalized objective over $\B$ using a gradient-based optimizer and updates the penalty parameters to encourage feasibility. Once $h(\B)$ is sufficiently small, a post-processing step such as thresholding tiny entries may be used to remove negligible edge weights.

\subsection{Statistical implications}
While replacing the combinatorial DAG constraint with a differentiable acyclicity constraint in \eqref{opt:notears_finite} removes a computational obstacle and makes gradient-based estimation feasible, it does not by itself resolve the statistical ambiguity inherent to observational data, as the objective may still admit multiple solutions with essentially the same fit. Recall that in our framework, sparsity plays an extra role beyond mere regularization, and distinguishes among observationally equivalent models by favoring those with fewer edges. This is precisely why we adopt the quasi-MCP penalty in \eqref{eq:quasi_MCP}---it more closely approximates $\ell_0$ edge counting than convex surrogates such as $\ell_1$, while remaining compatible with an end-to-end differentiable optimization framework. 

More broadly, differentiable DAG learning does not stop at being merely a computational device for avoiding a discrete search. Requiring a differentiable objective fundamentally shapes the estimator class and leads to new statistical questions, namely, (1) How and when can a GLM likelihood coupled with a sparsity penalty identify a special, meaningful causal structure (e.g. a minimal-edge $\mathcal{I}$-map) from observational data (cf. Theorem \ref{thm:identifiability})? (2) How can we analyze consistency of the finite sample optimization problem in \eqref{opt:notears_finite} over a continuous matrix space (e.g. with empirical process bounds)?, and (3) What are the nonasymptotic finite-sample guarantees for such an estimator (cf. Theorem \ref{thm.est.low})? These questions are an important step in further understanding the scope and inner-workings of differentiable DAG learning, and thus are addressed in the following section.

\section{Statistical guarantees}

The nonlinear program \eqref{opt:notears_finite} defines an estimator $\widehat B$ of the true DAG matrix $B^0$. In practice, it is known that these streamlined estimators have appealing properties. However, a detailed non-asymptotic characterization of their statistical properties is still largely missing from the picture. 

\subsection{Main result}

Our main theoretical result provides precisely such a characterization as we establish a high-probability bound for the event that $\widehat \pi \in \MinEqClassOrder$, i.e. that the learned causal order corresponds to a minimal-edge $\mathcal{I}$-map, and also establish finite-sample bounds on $\norm{\widehat B - B^*}_F^2$, the perturbation from that minimal representation. 

We now present the main finite sample result.

\begin{theorem}\label{thm.est.low}
Under \ref{assum.subex}--\ref{assum_glm}, let $s=1$ in case (a) and $s=3$ in case (b). Choose
\begin{align*}
 \lambda_0\asymp \sqrt{\frac{\log^s(np)}{n}},
 ~~\delta_0\asymp \sqrt{\frac{p^4\log^s(np)}{n}}
\end{align*}
with a fixed tuning constant for $\lambda_0$ and a sufficiently large fixed tuning constant for $\delta_0$. Assume $\tau\geq2\delta_0$ and $g_{\mathrm{gap}}\geq g_*>0$. Under $p^2\lesssim \sqrt{n/\log^{2s-1}(n)}$ %
with sufficiently small fixed tuning constants, with probability at least $1-2/(pn)$, global minimizers of \eqref{opt:notears_finite} exist and every global minimizer $\widehat B\sim\widehat\pi$ satisfies
\begin{align*}
 \widehat\pi\in\MinEqClassOrder,
 ~~~\min_{B^*\in\MinEqClass}\frac1p
 \norm{\widehat B-B^*}_F^2
 \lesssim \frac{p\log^s(np)}{n}
\end{align*}
\end{theorem}

Theorem \ref{thm.est.low} shows that every global minimizer selects an order corresponding to a minimal-edge $\mathcal I$-map. Its normalized squared coefficient error is of order $p/n$ up to log factors in both cases, reflecting the dense parameter space $B\in\mathbb R^{p\times p}$. Recall that this is a finite-sample guarantee for an estimator that is defined with a single differentiable penalized objective over \emph{all} DAG coefficient matrices across \emph{all} topological orders. 

We pause to examine two points that lead to challenges in finite-sample GLM-DAG theory: (i) Every variable $X$ is simultaneously an explanatory variable and a response variable, and (ii) We do not assume an apriori known topological order or skeleton, i.e. $\widehat \pi$ itself is random and data-dependent. 

The effect of (i) is deceptively far reaching in that unlike standard GLM analyses that assume random subexponential tails for the response variable and (WLOG) a deterministic boundedness assumption on the explanatory variables, our GLM-\emph{DAG} setting cannot allow such a mismatch, and instead requires mutually consistent regularity conditions under one joint distribution. Of course, this logic also applies to non-GLM, non-DAG MCP penalized regression settings, further distinguishing our problem from conventional ones. This trickiness is reflected in the regularity conditions \ref{assum.subex}-\ref{assum_glm} that separate the bounded setting and the unbounded setting. The bounded setting of \ref{assum.subex}(a), \ref{assum_glm}(a) leads to derivations and results more closely resembling those in the classical GLM literature, while the unbounded case of \ref{assum.subex}(b), \ref{assum_glm}(b) incorporates arguments that control both the truncated empirical fluctuations and the un-truncated population tails in a joint distribution analysis, leading to additional logarithmic terms in the final estimation rate. 

An unknown topological order from (ii) further complicates the analysis and requires uniform likelihood control across candidate orders, including those with misspecified working GLMs. In the remainder of the section, we provide a roadmap of the proof of Theorem~\ref{thm.est.low} with a focus on this uniform control.
The proof combines empirical process bounds, localization arguments, and separate score comparisons for compatible and incompatible orders. 

\subsection{Proof outline}

The first main ingredient in the proof is to control stochastic fluctuations over all orders and nodes. To uniformly control likelihood perturbations in local neighborhoods around the targets $\widetilde b^{\pi}_j$, we define for $v\in\mathbb R^{j-1}$ the ``Fisher norm''
\begin{align}
  \norm{v}_{\pi,j}^2 :=\frac{1}{\kappa}v^\top I_{\pi,j}v, \label{fishernorm}
\end{align}
which is a reweighted version of the Euclidean norm using the Fisher information matrix defined in \ref{assum_glm}. A crucial result is the following empirical process bound over all topological orders:

\begin{lemma}[Empirical process bound]\label{lem.emp.main}
Assume \ref{assum.subex}--\ref{assum_glm}, and let $s=1$ in case (a) and $s=3$ in case (b). Assume $p\lesssim \sqrt{n/\log^s n}$. For every deterministic $0<R\leq r_0$, with $r_0$ defined as
\begin{equation}\label{eq.r_0_main}
\begin{aligned}
 r_0:=\begin{cases}
 \displaystyle\frac1{\sqrt p}
 \min\left\{C_{\mathrm{reg}},\frac{\kappa}{2LC_{\max}}\right\},
 &\text{under case (a)},\\[6pt]
 \displaystyle\frac1{4C_{\max}\sqrt p\log(np)},
 &\text{under case (b)},
 \end{cases}
\end{aligned}
\end{equation}
where fixed constant $L>0$ satisfying $|A'''|<L$ exists from \ref{assum.subex}(a), \ref{assum_glm}(a), define
\begin{align*}
 \epsilon_n(R):=\sup_{\pi,j}
 \sup_{\norm{b-\widetilde b_j^\pi}_{\pi,j}\leq R}
 \left|
 \{\ell_{j,n}^\pi(b)-\ell_j^\pi(b)\}
 -\{\ell_{j,n}^\pi(\widetilde b_j^\pi)
       -\ell_j^\pi(\widetilde b_j^\pi)\}
 \right|.
\end{align*}
For a fixed constant $C'>0$, with probability at least $1-1/(pn)$,
\begin{align*}
 \epsilon_n(R)\leq C'R\sqrt{\frac{p\log^s(np)}{n}}.
\end{align*}
\end{lemma}

Lemma \ref{lem.emp.main} connects the population-level separation with its sample-level counterpart, which is evident through the following decomposition:
\begin{align}
    \ell^{\pi}_{j,n}(b) - \ell^{\pi}_{j,n}(\widetilde{b}^{\pi}_j)~=~\ell^{\pi}_{j}(b) - \ell^{\pi}_{j}(\widetilde{b}^{\pi}_j) +\bigg[\Big\{\ell^{\pi}_{j,n}(b) - \ell^{\pi}_j(b)\Big\} - \Big\{\ell^{\pi}_{j,n}(\widetilde{b}^{\pi}_j) - \ell^{\pi}_j(\widetilde{b}^{\pi}_j)\Big\}\bigg]. \label{eq.decomposition}
\end{align}
However, the connection with the finite-sample minimizer, $\widehat B$ is not yet sufficient as the proposed estimator is the minimizer of a \emph{penalized} likelihood objective \eqref{opt:notears_finite}, and also since the empirical process bound in Lemma \ref{lem.emp.main}, while controlling the likelihoods, does not on its own guarantee any direct bounds for the minimizer $\widehat B$ itself.

This leads to the second ingredient, which is a localization argument that is modified for a nonconvex objective. Standard convex localization arguments use positivity of the objective increment on the boundary of a ball to force every global minimizer to lie inside the ball \citep{negahban2012unified}. Recall however that $L_n$ in \eqref{eq:sample_obj} is nonconvex due to the quasi-MCP in \eqref{eq:quasi_MCP}. This penalty related issue is resolved by exploiting its $\ell_1$-Lipschitz property, which leads to a modified localization argument that shows that deriving a local estimation error bound reduces to showing that a certain constructed function (shown in \eqref{eq.localization}) is positive on the boundary of a ball:

\begin{equation}\label{eq.localization}
\begin{aligned}
 \inf_{\norm{b-\widetilde b_j^\pi}_{\pi,j}=R}
 \left\{
 \ell_{j,n}^\pi(b)-\ell_{j,n}^\pi(\widetilde b_j^\pi)
 -\lambda_0\norm{b-\widetilde b_j^\pi}_1
 \right\}>0.
\end{aligned}
\end{equation}

This also elucidates a key reason why empirical process bounds are used for the proof. Localization arguments require uniform control on all candidate $b$ on the boundary of a ball, and an empirical process bound provides exactly this by uniformly controlling perturbations on (and within) a ball. Thus, the upper bound in Lemma \ref{lem.emp.main} can be directly plugged in to \eqref{eq.localization} to further lower bound the infimum and establish positivity.

There is one final step: While localization is key in deriving a concrete finite-sample estimation error bound for each topological order, it alone does not compare the candidates across \textit{all} orders. Thus, the final step is establishing the following comparison between fitted empirical objectives: 
\begin{align}
    \underset{\substack{\pi_0 \in \MinEqClassOrder\\ \pi_2 \not\in \MinEqClassOrder}}{\min}~L_n\big(\widehat{B}(\pi_2);\lambda_0,\delta_0\big)-L_n\big(\widehat{B}(\pi_0);\lambda_0,\delta_0\big)~>~0, \label{eq.global.comparison}
\end{align}
where for each topological order $\pi$, $\widehat B(\pi) \in\arg\min_{B \sim \pi} L_n(B; X, \lambda_0, \delta_0)$ denotes an order-specific empirical minimizer. For compatible non-minimal orders, each additional edge incurs a saturated MCP penalty of $\lambda_0\delta_0/2$, and choosing an appropriate constant for $\delta_0$ makes this penalty difference dominate likelihood fluctuations in $\widehat B(\pi_2), \widehat{B}(\pi_0)$. For incompatible orders, the theorem's separation conditions ensure that the population likelihood gap $g_{\mathrm{gap}}$ dominates potential penalty advantages and empirical fluctuations. This establishes \eqref{eq.global.comparison}, which implies $\widehat\pi\in\MinEqClassOrder$, while applying localization once more implies the final Frobenius norm error rate.

\section{Experiments}\label{sec.experiments}

Owing to the differentiability\footnote{Recall that due to the MCP penalty, our method is technically only subdifferentiable, however, standard gradient-based nonsmooth optimization schemes apply.}  of our method, the proposed objective is compatible with modern automatic-differentiation frameworks, making it straightforward to implement with standard gradient-based optimization tools. Here, we use a simpler implementation to focus on the statistical behavior of the method and alignment with the theory. We evaluate it across different GLM families and tasks, measuring its sample complexity and accuracy compared to several baselines including a negative control.

\subsection{Algorithm and implementation}

\begin{algorithm}[t]
\DontPrintSemicolon 
\SetAlgoLined
\LinesNumbered
\caption{\myalg{}}
\label{alg:dagma}
\KwIn{Data matrix $\bfX$, initial loss weight $\mu^{(0)}$, decay factor $\gamma_{\mu} \in (0,1)$, quasi-MCP parameters $\delta,\lambda$, number of iterations $T$, log-det parameters $s=(s^{(0)},\ldots,s^{(T-1)})\in(0,\infty)^T$, }
\KwOut{$B^{(T)}$}
initial matrix $\B^{(0)}$ (for example,  $\mathbf{0}\in\R^{\p\times \p}$)\;
\For{$t=0,1,2,\ldots,T-1$}{
  Starting from $\B^{(t)}$, solve $B^{(t+1)} = \arg\min_{\B\in \R^{p\times p}} \mu^{(t)}L_n(\B;\bfX,\lambda,\delta)+h(\B;s^{(t)})$\;
  Set $\mu^{(t+1)} = \gamma_\mu\, \mu^{(t)}$\;
}
\end{algorithm}
  
To solve \eqref{opt:notears_finite}, we primarily adopt \textsc{DAGMA} \citep{bello2022dagma}, a method that incorporates a differentiable acyclicity constraint into the objective, applies a gradient-based optimizer to the resulting unconstrained problem, and then gradually strengthens the acyclicity enforcement by increasing a penalty parameter while warm-starting from the previous solution. A general outline of the method, called \myalg{}, is presented in Algorithm \ref{alg:dagma}. 

\begin{figure}[t]
    \centering
    \captionsetup{font=footnotesize}
    \captionsetup[subfigure]{font=small}
    \begin{subfigure}[t]{0.49\linewidth}
        \centering
        \includegraphics[width=\linewidth]{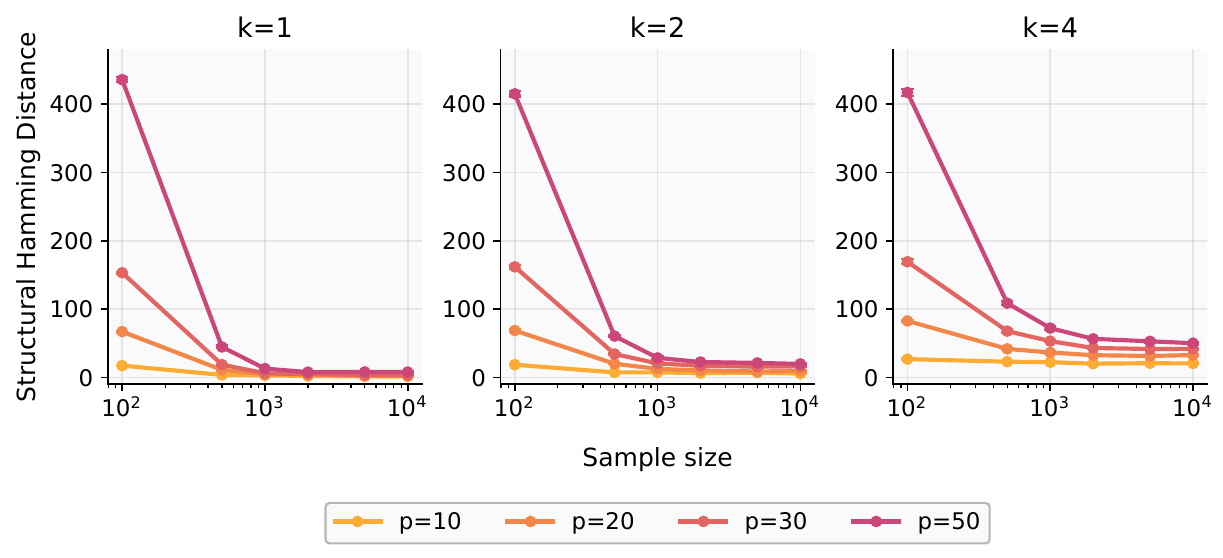}
        \caption{ER graph}
    \end{subfigure}
    \hfill
    \begin{subfigure}[t]{0.49\linewidth}
        \centering
        \includegraphics[width=\linewidth]{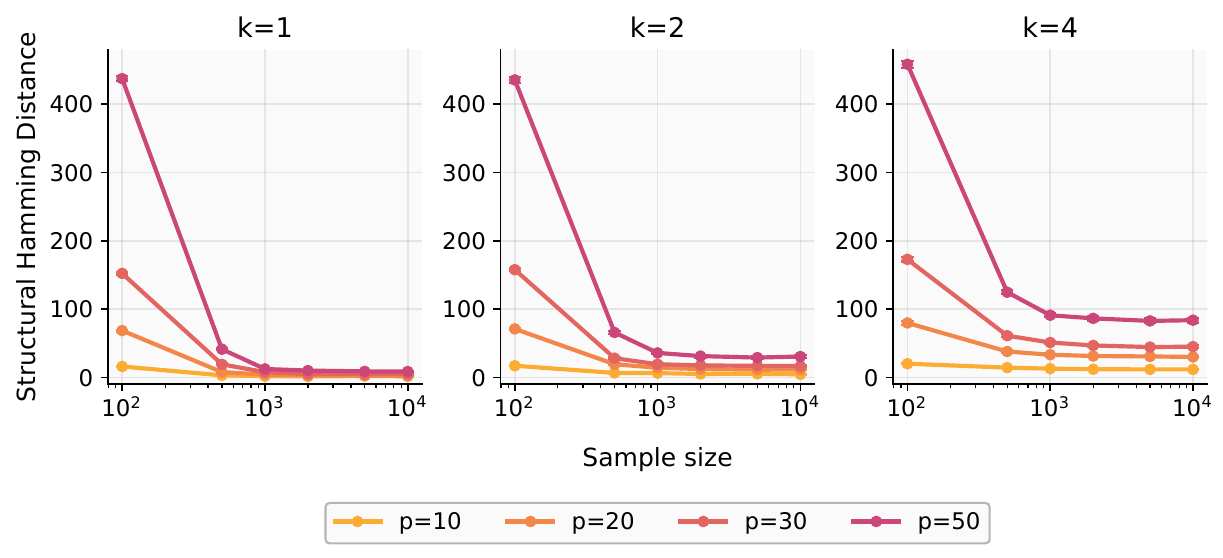}
        \caption{SF graph}
    \end{subfigure}
    \caption{Sample-size effect for the \textbf{logistic} model. SHD (lower is better) is plotted against sample size $n\in\{100,500,1000,2000,5000,10000\}$ for several graph sizes $p$. The left panel corresponds to Erd\H{o}s--R\'enyi (ER) graphs and the right panel corresponds to scale-free (SF) graphs.} 
    \label{fig:sample_size_two_panels_logistic}
\end{figure}

\subsection{Setup}
We consider two nonlinear GLMs (logistic for binary data and Poisson for count data), along with the classical linear Gaussian model as a benchmark for comparisons in Section \ref{subsubsec:RecoveryAccuarcy}. Unless otherwise stated, we generate the weighted adjacency matrix $B$ by first sampling a random DAG from either an Erd\H{o}s--R\'enyi (ER) family or a scale-free (SF) family, with density parameter $k\in\{1,2,4\}$. Given the sampled edge set $E$, we set $B_{i,j}=0$ for $(i,j)\notin E$, and for each $(i,j)\in E$ we draw an independent edge weight:
\begin{align}\label{eq.generation_B}
  B_{i,j} \sim \mathrm{Uniform}([-b_{\mathrm{upper} },-b_{\mathrm{lower}}]\cup [b_{\mathrm{lower}},b_{\mathrm{upper}}]),
\end{align}
where $0<b_{\mathrm{lower}}<b_{\mathrm{upper}}$. 
For the Gaussian benchmark, we use a linear Gaussian internally standardized structural causal model \citep[iSCM,][]{ormaniec2025standardizing} and apply columnwise sample standardization before fitting. By construction, iSCMs avoid known pitfalls of generating linear SCMs 
\citep{reisach2021beware}.

For the logistic and Poisson SEMs, we follow the formulation in (\ref{eq.logistic.1}-\ref{eq:logistic_nll}) and (\ref{eq.poisson.1}-\ref{eq:poisson_nll}), respectively, with $B$ generated as in \eqref{eq.generation_B} for both models.
For the Gaussian model, we use the formulation in (\ref{eq.Gaussian.1}--\ref{eq:gaussian_nll}), with coefficients and noise variances determined by the iSCM construction.

\subsection{Results}

\subsubsection{Sample complexity}

We first examine how the sample size affects structure recovery across each setting using the Structural Hamming Distance (SHD) as the evaluation metric. We report the SHD computed at the level of completed partially directed acyclic graphs (CPDAGs), i.e. we convert each estimated graph to its CPDAG and compare it with the CPDAG of the data-generating graph.

\begin{figure}[t]
    \centering
    \captionsetup{font=footnotesize}
    \captionsetup[subfigure]{font=small}
    \begin{subfigure}[t]{0.49\linewidth}
        \centering
        \includegraphics[width=\linewidth]{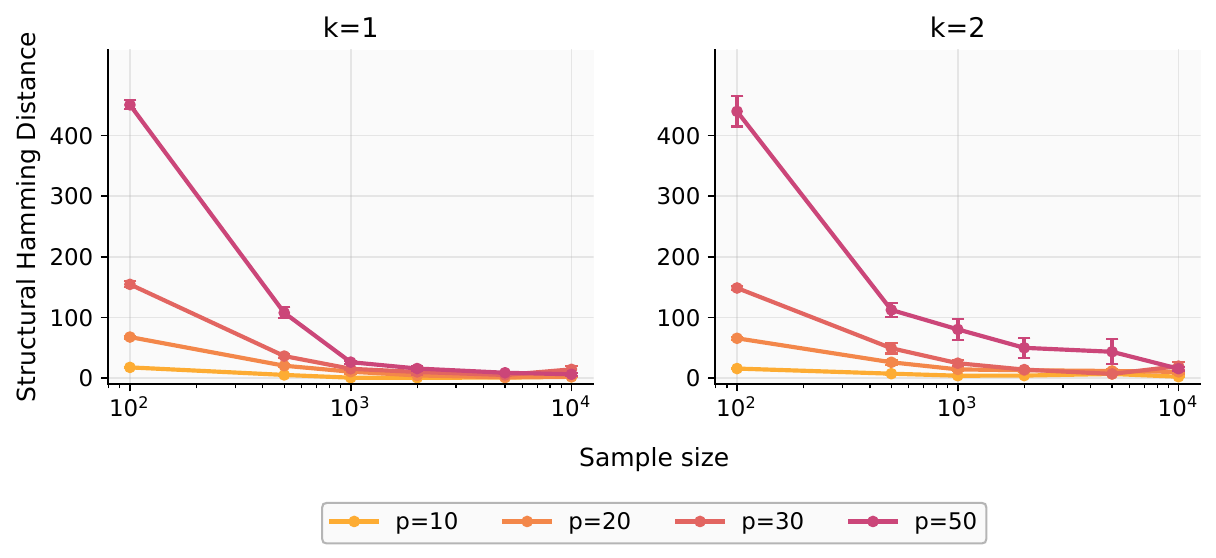}
        \caption{Poisson ER}
    \end{subfigure}
    \hfill
    \begin{subfigure}[t]{0.49\linewidth}
        \centering
        \includegraphics[width=\linewidth]{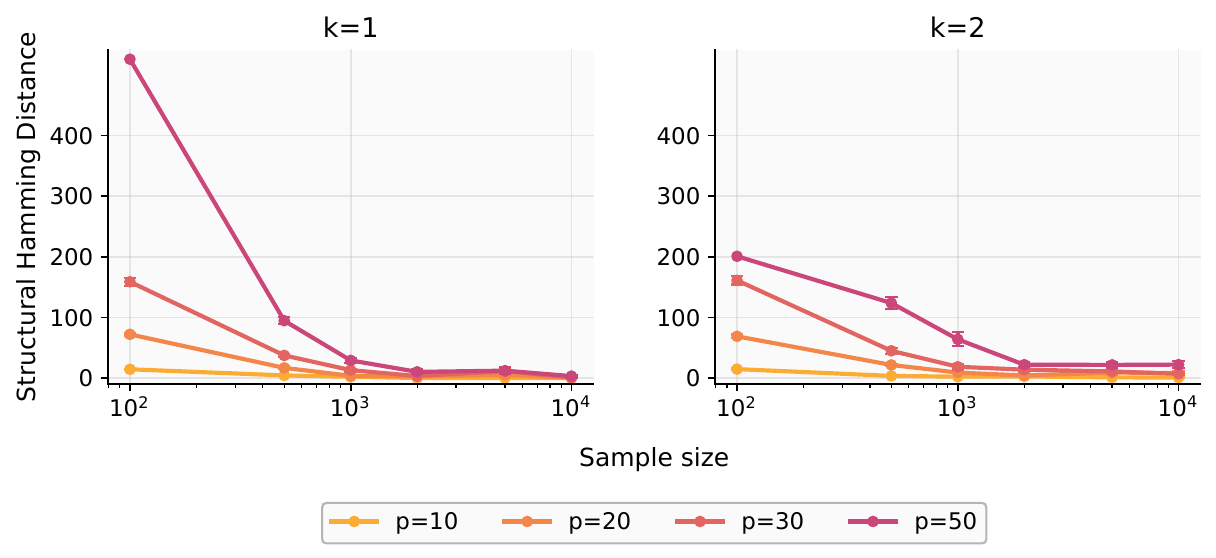}
        \caption{Poisson SF}
    \end{subfigure}
    \caption{Sample-size effect for the \textbf{Poisson} model. SHD (lower is better) is plotted against sample size $n\in\{100,500,1000,2000,5000,10000\}$ for several graph sizes $p$. The left panel corresponds to Erd\H{o}s--R\'enyi (ER) graphs and the right panel corresponds to scale-free (SF) graphs.}
    \label{fig:sample_size_two_panels_poisson}
\end{figure}

Figures~\ref{fig:sample_size_two_panels_logistic} and \ref{fig:sample_size_two_panels_poisson} plot the SHD as a function of $n$ for several graph sizes $p$ and for Erd\H{o}s--R\'enyi and scale-free graphs in the logistic and Poisson settings, respectively. The overall pattern is consistent across all model classes in that as the sample size increases, the SHD decreases.

The plots show that sample requirements grow with graph size. Larger graphs require more observations to attain the same SHD level as smaller graphs. This pattern is visible across all model classes and both ER and SF graphs. Overall, these experiments provide empirical verification of theoretical statistical consistency.

\subsubsection{Negative control}

\begin{table}[t]
\centering
\caption{Negative Controls for \myalg{}.}
\label{table:negative_controls}
\small
\setlength{\tabcolsep}{6pt}
\renewcommand{\arraystretch}{1.15}

\begin{tabular}{@{}llrcrcr@{}}
\toprule
\multirow{2}{*}{Model}
& \multirow{2}{*}{Graph type}
& \multicolumn{2}{c}{\textbf{\myalg{}}}
& \multicolumn{2}{c}{\textbf{Negative control}}
& \multirow{2}{*}{$\hat{q}_{\text{NC}}$} \\
\cmidrule(lr){3-4}
\cmidrule(lr){5-6}
& & Mean & 95\% & Mean & 95\% & \\
\midrule

\multirow{4}{*}{Logistic}
& ER-1 & 2.311 & (0.0, 9.0)
& 16.853 & (13.0, 20.0) & 0.0 \\

& ER-4 & 16.735 & (8.0, 26.0)
& 33.47 & (20.975, 37.0) & 0.028 \\

\cmidrule(l){2-7}

& SF-1 & 1.727 & (0.0, 6.0)
& 15.165 & (12.0, 19.0) & 0.0 \\

& SF-4 & 12.232 & (4.0, 22.0)
& 29.928 & (23.0, 38.0) & 0.003 \\

\midrule

\multirow{4}{*}{Poisson}
& ER-1 & 2.737 & (0, 12)
& 17.878 & (14, 22) & 0.007 \\

& ER-3 & 7.621 & (0, 22)
& 32.031 & (24, 39) & 0.003 \\

\cmidrule(l){2-7}

& SF-1 & 1.261 & (0, 7.025)
& 16.041 & (12, 19) & 0.004 \\

& SF-3 & 4.14 & (0, 18)
& 29.956 & (23, 36) & 0.001 \\

\bottomrule
\end{tabular}
\vspace{5pt}
\begin{minipage}{\linewidth}
\setstretch{1}
\footnotesize
95\% indicates the central 95\% interval (i.e. 2.5\%-97.5\% percentiles) of SHD across simulation replicates. $\hat{q}_{\text{NC}}$ is the fraction of replicates in which the negative control attains an SHD no larger than that of the evaluated method.
\end{minipage}
\end{table}

To assess whether the performance of \myalg{} reflects genuine recovery of causal structure rather than random guessing, we additionally use \emph{negative controls} \citep{petersen2024you}. The main idea is to compare the estimated CPDAG with a \emph{randomly generated} CPDAG of comparable complexity, as measured by the number of edges, but otherwise unrelated to the data-generating graph. Concretely, for each simulation replicate, we record the number of edges in the final graph estimated by \myalg{} and generate a random DAG on the same nodes with exactly that many edges. We compute the SHD for both \myalg{} and the negative control, and we report the empirical comparison frequency $\hat{q}_{\text{NC}}$, defined as the fraction of replicates in which the negative control attains an SHD no larger than \myalg{}. This baseline controls for the fact that some metrics can be artificially improved by outputting graphs that are systematically too sparse or too dense, and directly assesses whether \myalg{} performs better than random guessing at the same edge density.

For the Poisson negative-control experiments, we use $k=3$ instead of $k=4$ to avoid numerical overflow. Table~\ref{table:negative_controls} summarizes the results for each model class at $p=10$ across graph families, based on $1000$ repetitions. Across both ER and SF graphs and across the sparsity levels considered, \myalg{} substantially outperforms the corresponding negative controls.

\subsubsection{Benchmark comparisons}\label{subsubsec:RecoveryAccuarcy}

Finally, we provide detailed comparisons against prior methods for causal structure learning. We start by mentioning a few comments for each setting:
\begin{itemize}
    \item Recent methods are generally tailored to the linear Gaussian setting and are not directly applicable to nonlinear models such as GLMs. Although we include the Gaussian case for comparison, we emphasize that  the focus of our study is nonlinear causal structure learning, and accordingly our method is not optimized for the linear Gaussian setting.
    \item The exceptions to the above are classical general purpose methods such as PC and (F)GES that can be applied to discrete data as well. Thus, PC and FGES are used as benchmarks in the logistic setting.
\end{itemize}
\noindent
In addition to SHD, we record computational runtime to assess scalability.

\paragraph{Gaussian model.}
In the Gaussian setting, we run comparisons on standardized data using iSCM \mbox{\citep{ormaniec2025standardizing}} followed by sample standardization. The experiment uses \mbox{$n=1000$}, \mbox{$p\in\{10,20,30,50,80,100\}$}, ER and SF graphs, \mbox{$k\in\{1,2,4\}$} per setting. Each dataset is shared by all methods evaluated in that setting. We evaluate \myalg{} against standard baselines including PC \citep{spirtes2000causation}, FGES \citep{ramsey2017million}, and MICODAG-CD (coordinate descent) \citep{xu2025asymptotically}. For MICODAG-CD, we provide the moral graph as additional input, which can substantially improve its performance. \myalg{} does not use this side information.

\begin{figure}[t]
    \centering
    \captionsetup{font=footnotesize}
    \captionsetup[subfigure]{font=small}
    
    \begin{subfigure}[t]{0.49\linewidth}
        \centering
        \includegraphics[width=\linewidth]{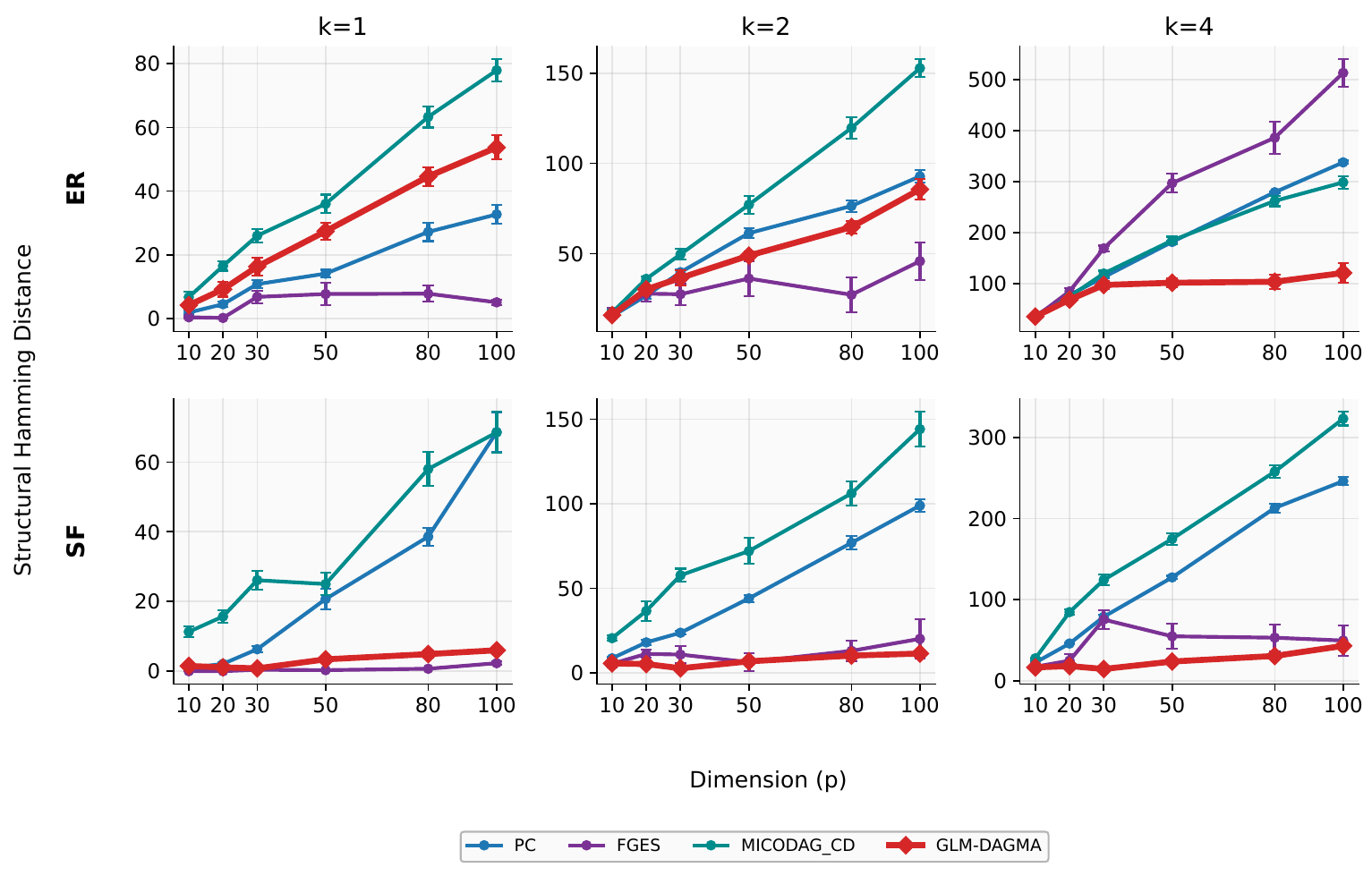}
        \caption{SHD to the true MEC (lower is better).}
        \label{fig:gauss_std_shd}
    \end{subfigure}
    \begin{subfigure}[t]{0.49\linewidth}
        \centering
        \includegraphics[width=\linewidth]{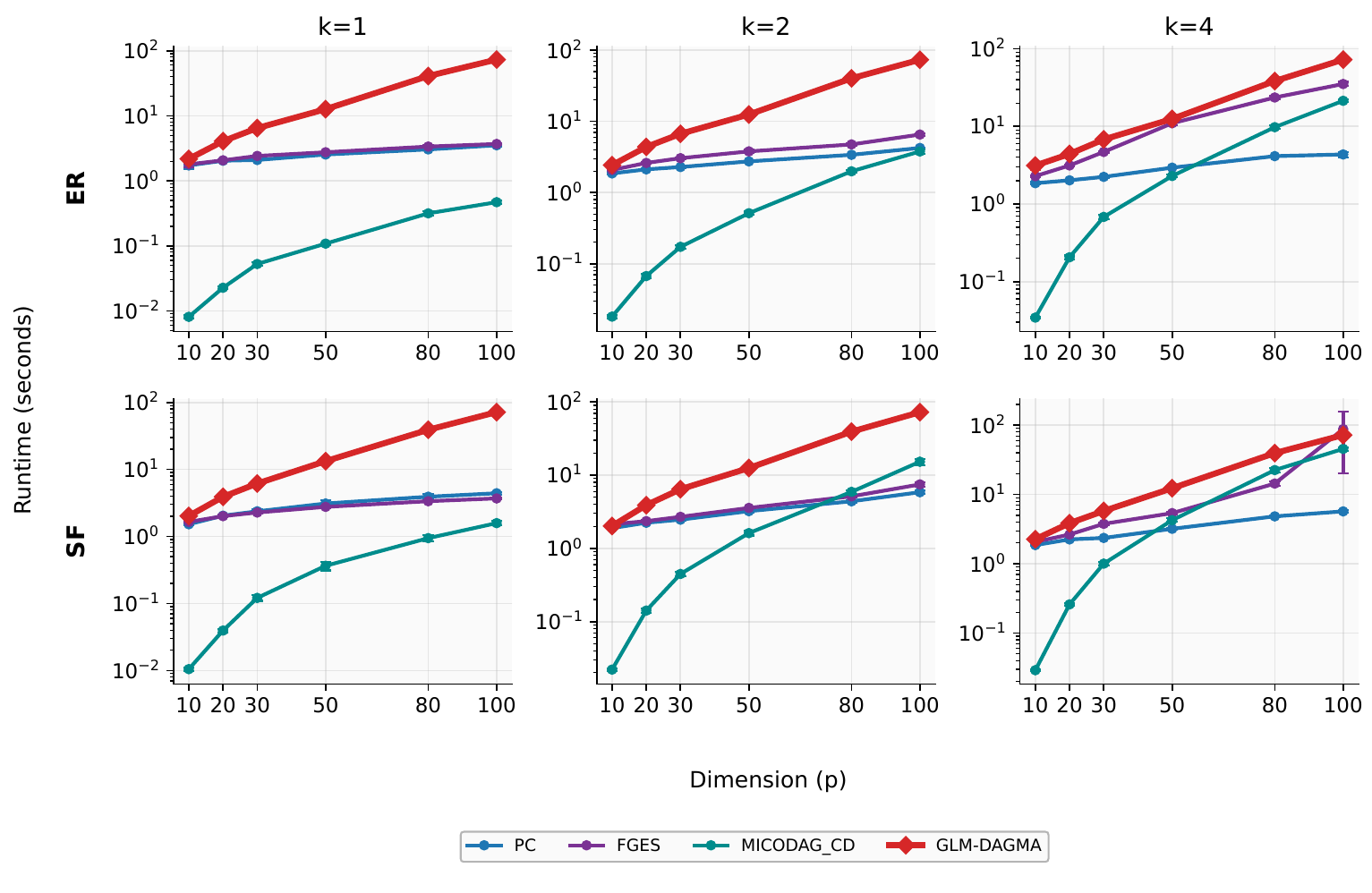}
        \caption{Runtime (log scale; lower is better).}
        \label{fig:gauss_std_time}
    \end{subfigure}
    \caption{Standardized Linear Gaussian model results with $n=1000$. \textbf{Left:} SHD between the estimated graph and the ground truth. \textbf{Right:} runtime (log scale). Columns correspond to graph density $k\in\{1,2,4\}$ and rows correspond to graph families $\{\mathrm{ER},\mathrm{SF}\}$. Graph size $p\in\{10,20,30,50,80,100\}$ is plotted on the $x$-axis.} 
     \label{fig:gauss_std_shd_time}
\end{figure}

The SHD-runtime results 
are shown in Figure~\ref{fig:gauss_std_shd_time}. Many competing methods struggle to recover a meaningful graph structure. Overall, GLM-DAGMA provides a strong time-accuracy tradeoff across settings as it achieves the best overall SHD performance while remaining practical in runtime. The gains are most pronounced on larger and denser graphs, where GLM-DAGMA remains robust and finishes in only a couple minutes (e.g., $p=100$, $k=4$).

\begin{figure}[t]
    \centering
    \captionsetup{font=footnotesize}
    \captionsetup[subfigure]{font=small}
    \begin{subfigure}[t]{0.49\linewidth}
        \centering
        \includegraphics[width=\linewidth]{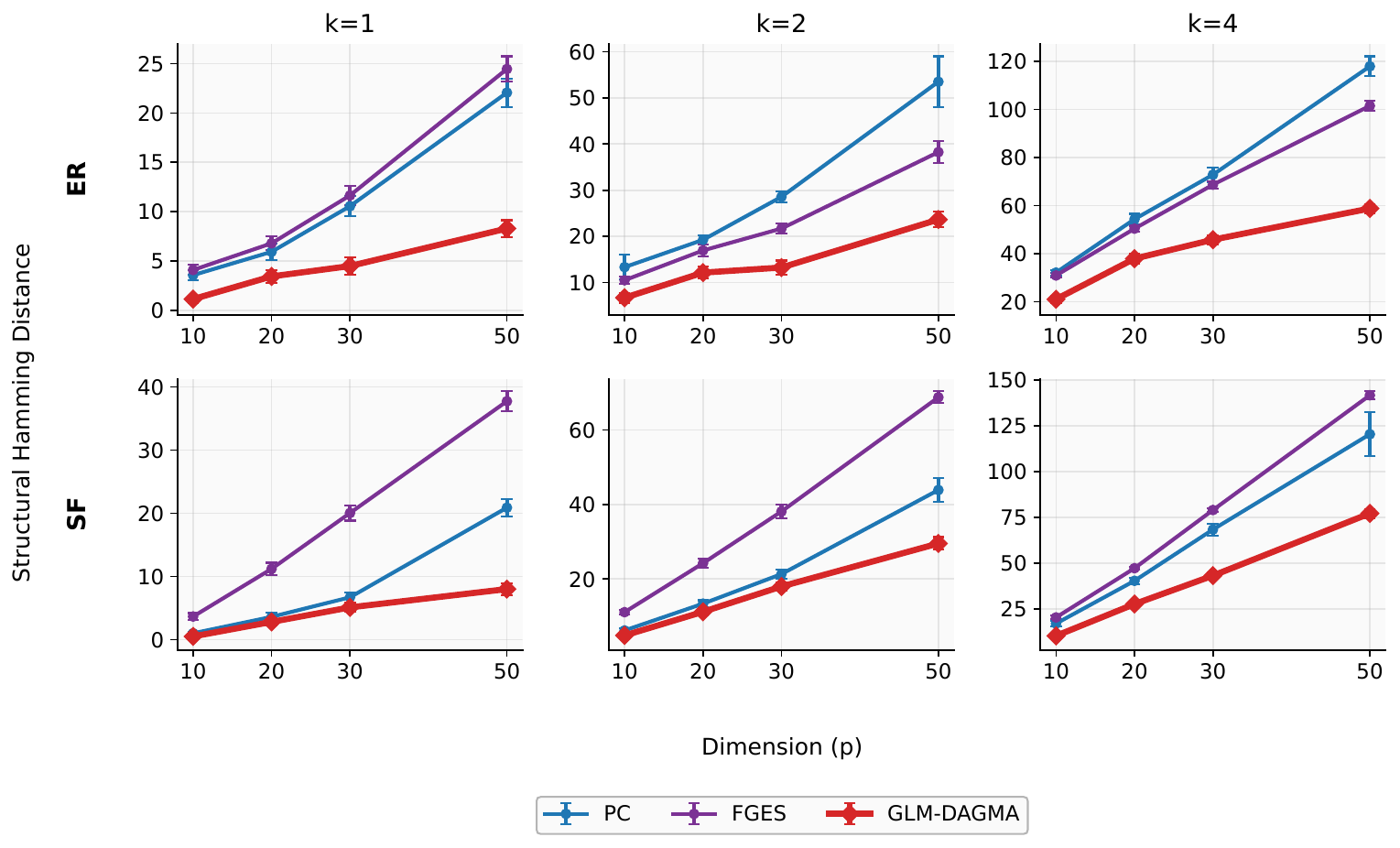}
        \caption{SHD to the true MEC (lower is better).}
        \label{fig:logistic_shd}
    \end{subfigure}
    \begin{subfigure}[t]{0.49\linewidth}
        \centering
        \includegraphics[width=\linewidth]{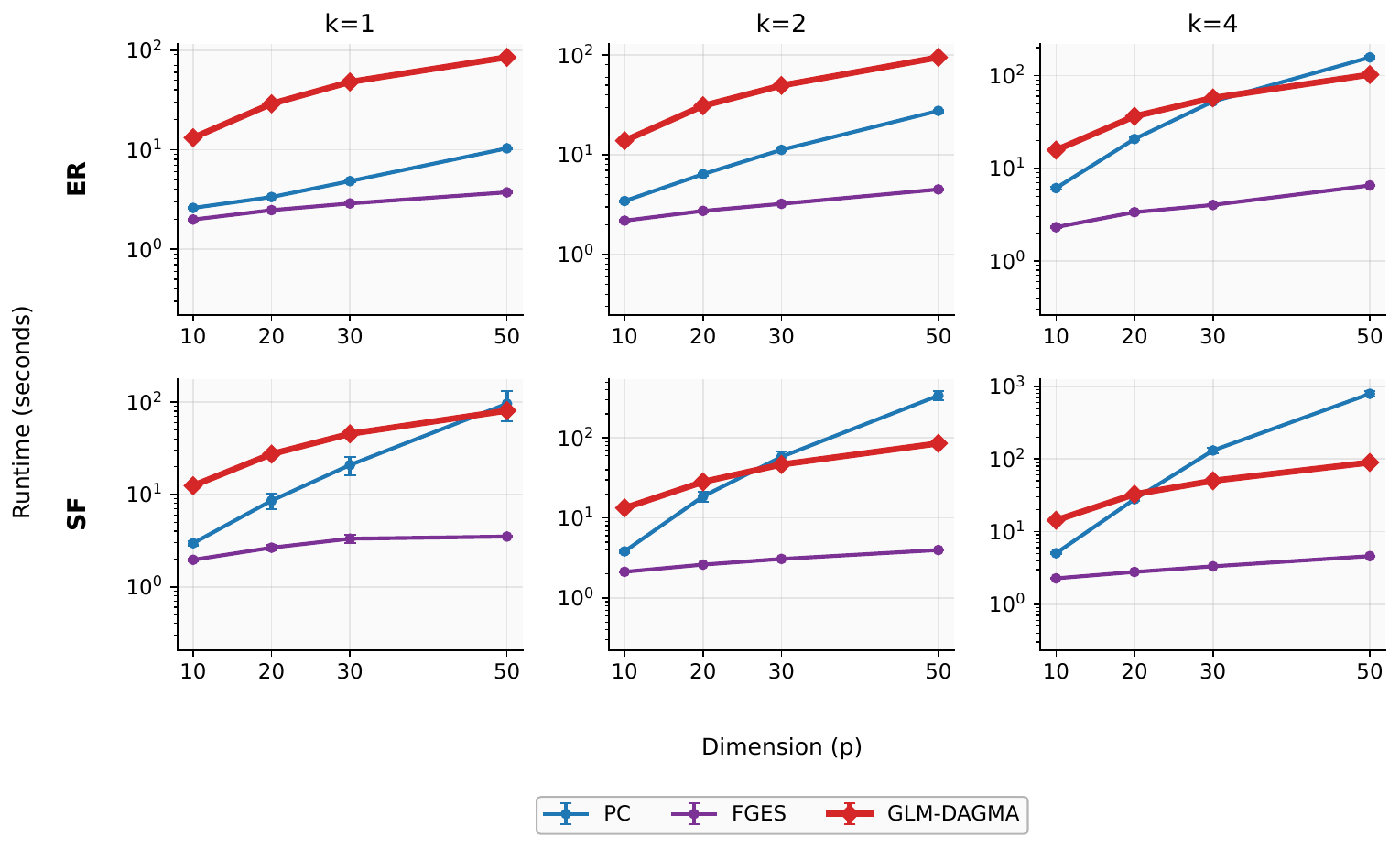}
        \caption{Runtime (log scale; lower is better).}
        \label{fig:logistic_time}
    \end{subfigure}
    \caption{Logistic model results with sample size $n=5000$. \textbf{Left:} SHD between the estimated graph and the ground truth. \textbf{Right:} runtime (log scale). Columns correspond to graph density $k\in\{1,2,4\}$ and rows correspond to graph families $\{\mathrm{ER},\mathrm{SF}\}$. Graph size $p\in\{10,20,30,50\}$ is plotted on the $x$-axis.}
    \label{fig:logistic_shd_time}
\end{figure}

\paragraph{Logistic model.}
The SHD-runtime results for logistic SEMs are shown in Figure~\ref{fig:logistic_shd_time}. We compare \myalg{} against PC and FGES. \myalg{} delivers the best overall SHD performance across graph families and densities, and its advantage becomes increasingly pronounced as the graphs grow larger and denser; see Figure~\ref{fig:logistic_shd}. The runtime is driven mainly by graph size and shows little sensitivity to graph family or density. 

\begin{figure}[t]
    \centering
    \captionsetup{font=footnotesize}
    \captionsetup[subfigure]{font=small}
    \begin{subfigure}[t]{0.49\linewidth}
        \centering
        \includegraphics[width=\linewidth]{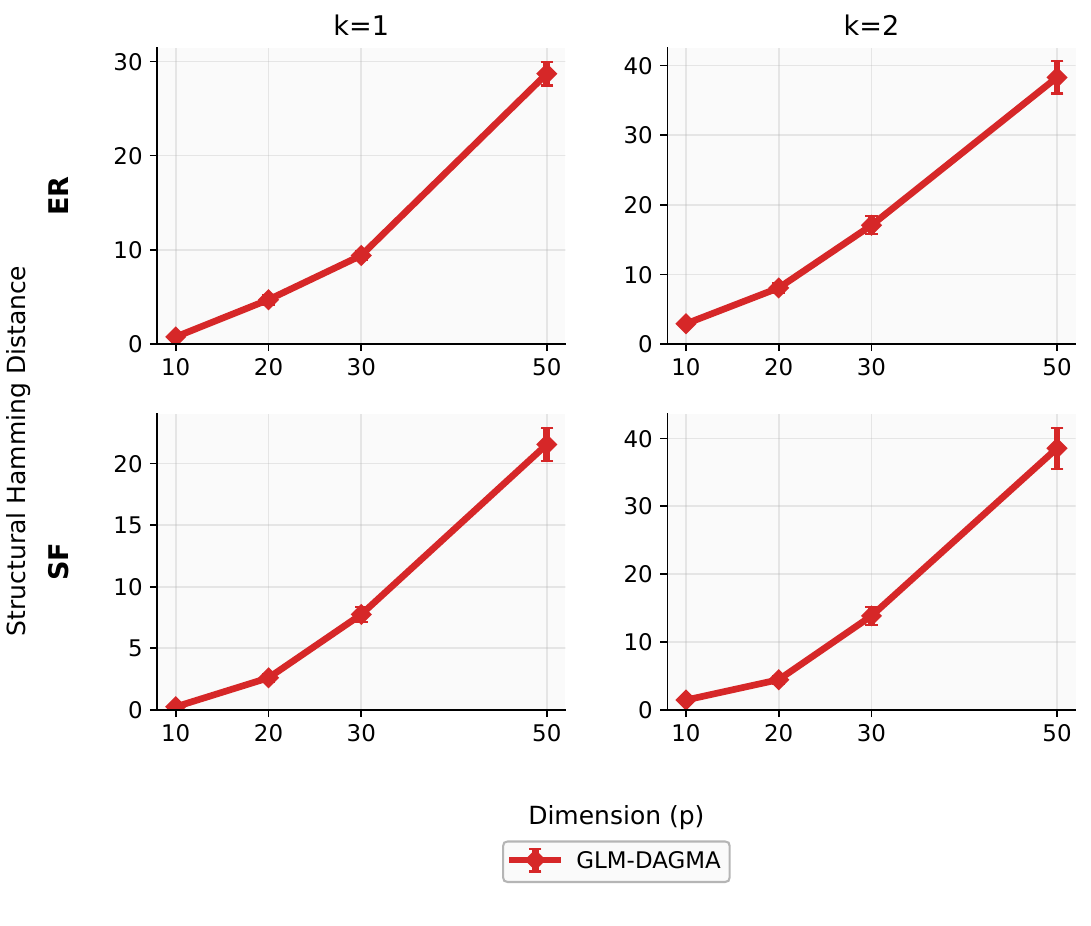}
        \caption{SHD to the true MEC (lower is better).}
        \label{fig:poisson_shd}
    \end{subfigure}
    \hfill
    \begin{subfigure}[t]{0.49\linewidth}
        \centering
        \includegraphics[width=\linewidth]{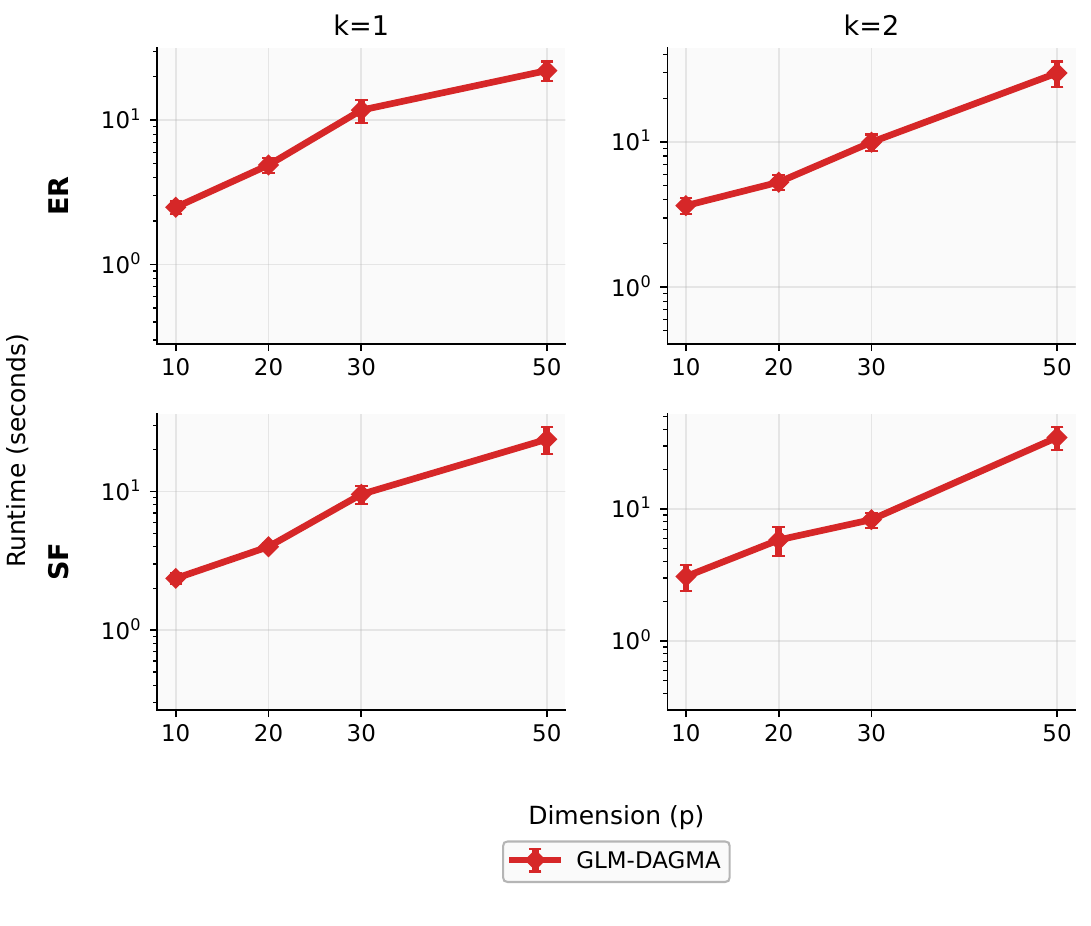}
        \caption{Runtime (log scale; lower is better).}
        \label{fig:poisson_time}
    \end{subfigure}
    \caption{Poisson model results with sample size $n=1000$. \textbf{Left:} SHD between the estimated graph and the ground truth. \textbf{Right:} runtime (log scale). Columns correspond to graph density $k\in\{1,2\}$ and rows correspond to graph families $\{\mathrm{ER},\mathrm{SF}\}$. Graph size $p\in\{10,20,30,50\}$ is plotted on the $x$-axis.}
    \label{fig:poisson_shd_time}
\end{figure}

\paragraph{Poisson model.}
The SHD-runtime results for Poisson SEMs are shown in Figure~\ref{fig:poisson_shd_time}. This setting is numerically more challenging because the Poisson means $\exp(\mathbf{x}_{\cdot i}^\top B_{\cdot j})$ (see \eqref{eq:poisson_nll}) can become large, triggering computational difficulties. Thus, we focus on the $k\in\{1,2\}$ setting.
To the best of our knowledge, open-source implementations of existing methods are not applicable or unavailable in this setting, so we report results only for \myalg{}. The results show \myalg{} consistently recovers a nontrivial graph structure, with an SHD substantially better than that of the empty graph baseline across the settings considered. The runtime results also remain practical across all configurations.

\section{Discussion}
We have studied differentiable causal structure learning with canonical conditional GLMs through a single objective combining a negative log-likelihood, an exact acyclicity constraint, and a quasi-MCP penalty. Under the stated conditions, population minimizers are minimal-edge representations, and empirical minimizers satisfy finite-sample statistical guarantees. Empirical experiments illustrate the utility of differentiable methods, which are easily implemented using standard gradient-based solvers, and either meet or exceed the performance of existing baselines. We hope this provides a unified framework for studying identifiability, estimation, and computation across this broad class of models.

\bibliographystyle{abbrvnat}
\bibliography{ref}

\end{document}